# Phonon Thermal Transport between Two in-Plane, Two-Dimensional Nanoribbons in the Extreme Near-Field Regime

Md Jahid Hasan Sagor[1] and Sheila Edalatpour[1,2]

[1] Department of Mechanical Engineering, University of Maine, Orono, Maine 04469, USA

[2] Frontier Institute for Research in Sensor Technologies, University of Maine, Orono, Maine 04469, USA

**Abstract**

The phonon thermal conductance of sub-nanometric vacuum gaps between two in-plane nanoribbons of two-dimensional materials (graphene and silicene) is analyzed using the atomistic Green's function method and by employing the Tersoff and Lennard-Jones potentials for describing the interatomic interactions. It is found that the phonon conductance decay exponentially with the size of the gap. Three exponential regimes have been identified. In the regime where the Lennard-Jones (L-J) potential is driven by the repulsive interatomic forces, caused by the overlap of electronic orbits, there is a sharp exponential decay in conductance as the gap increases ($e^{-10.0d}$ for graphene). When both the repulsive and attractive (van der Waals) interatomic forces contribute to the L-J potential, the decay rate of the conductance significantly reduces to $e^{-2.0d}$ for graphene and $e^{-2.5d}$ for silicene. In the regime where attractive van der Waals forces dominate the L-J potential, phonon conductance has the slowest exponential decay as $e^{-1.3d}$ for both silicene and graphene. It is also found that the contribution from the optical phonons to the conductance is non-negligible only for very small gaps between graphene nanoribbons ($d <$ 1.6 Å). The phonon conductance of the gap is shown to vary with the width of the nanoribbon very modestly, such that the thermal conductivity of the gap linearly increases with the nanoribbon widths. The results of this study are of significance for fundamental understanding of heat transfer in the extreme near-field regime and for predicting the effect of interfaces and defects on heat transfer.

## I. Introduction

Heat transfer between two media separated by a vacuum gap is said to be in the extreme near-field regime when the gap size is smaller than a few nanometers. Heat transfer in this regime is of significance both from a fundamental point of view and for applications in heat-assisted magnetic recording [1,2], scattering tunneling microscopy [3], and interfacial heat transfer, to name only a few. It is now known that heat transfer in the extreme near-field regime can simultaneously occur by phonons, photons, and electrons (in the case of metallic media) [4-25]. Recent studies comparing the contributions from these three energy carriers to the total thermal conductance in the extreme near-field regime have found that, in the absence of a bias voltage, phonons are the dominant heat transfer mechanism at sub-nanometric separation gaps [4-6,10,14,15,18,20-22]. So far, phonon heat transfer through a vacuum gap has been studied using the nonequilibrium Green's function method [6,8,12,18,20-22], non-equilibrium molecular dynamics simulations [19,23,25], elastic continuum model [4,9,10,13], scattering boundary method [7], lattice dynamics [5,15], and fluctuational electrodynamics [14,16,24]. These studies have provided significant insight into phonon heat transfer through an extremely small vacuum gap between dielectric [6-8,10,15,23,25], semiconducting [4,5,20] and metallic [9,11,13,14,16-19,21,22,24] media. It has been found that the contribution of phonons to the heat flux cannot be neglected, and energy transfer by phonons through the gap significantly enhances thermal conductance [4-6,9,14,19-24]. Non-equilibrium molecular dynamic simulations have shown that phonon conductance exponentially decreases with increasing the separation gap between two planar platinum slabs [19]. The atomistic Green's function method used with the density functional theory simulations have shown a gap-dependent conductance as $d^{-11.9 \pm 1.2}$ for two silicon planar media [20]. Using the nonequilibrium Green's function method and molecular dynamics simulations, it has been revealed that anharmonicity only has a moderate effect on the phonon conductance of a gap between two metallic media [22], while phonon conductance for a gap between dielectric media is significantly affected by the anharmonic effects [23]. The effect of a bias voltage, which is applied in the experimental studies on vacuum gaps between metallic media [3, 21, 26-29], on the conductance of the gap has also been studied. A recent study has found that the low bias voltage applied in experimental studies has a weak effect on the phonon heat transfer [22]. Some studies have shown that the main carrier responsible for transferring heat through the gap depends on the applied bias voltage, and the contribution from the electrons to the heat flux is dominant for the voltages in the same order of those applied in the experimental studies [18,24]. The thermal conductance through a gap mediated by the phonons induced by the Casimir force has also been analyzed, and it is shown that the contribution of these phonons to thermal transport is non-negligible when the gap size is smaller than the lattice constant [7].

While phonon conductance through extremely small vacuum gaps has been studied for dielectric, semiconducting, and metallic media, no study has been dedicated to phonon conductance of a gap separating two-dimensional materials. The phonon thermal transport within single (e.g., Ref. [30]) and multiple layers (e.g., Ref. [31]) of two-dimensional materials has also been studied, but no study is done on phonon conductance across a vacuum gap separating two two-dimensional materials. In this paper, we study phonon heat transfer between two in-plane, semi-infinite nanoribbons of graphene and silicene separated by a sub-nanometric vacuum gap. Due to superior thermal properties, graphene has found several applications in thermal management [30], energy storage [32], and electronics [33,34]. The supreme electronic properties of silicene combined with its low thermal conductivity make this two-dimensional material a promising candidate for thermoelectric devices [35]. Additionally, silicene can be exploited for applications in nanoelectronics, particularly since it can more easily be integrated in silicon-based electronics compared to graphene [36-38]. Understanding heat transfer through a sub-nanometric gap between these two-dimensional materials is important not only from a fundamental point of view, but also for predicting thermal resistance in the presence of cracks and nanovoids in these materials [39]. We estimate the phonon conductance of the gap using the atomistic Green's function method and by utilizing the Tersoff and Lennard-Jones potentials for modeling the interatomic interactions in the nanoribbons and across the gap, respectively. We analyze how the conductance of the gap varies with the size of the gap as well as the temperature and width of the nanoribbons. We demonstrate exponential laws for phonon conductance versus gap size and compare the contributions from the acoustic and optical phonons to the thermal conductance.

The rest of this paper is organized as follows. The problem under study is explained in Section II, and the methods employed for solving this problem are presented in Section III. The results are provided in Section IV, and the concluding remarks are summarized in Section V.

## II. Description of the Problem

The problem under study is schematically shown in Fig. 1a. Two in-plane, semi-infinite nanoribbons of a two-dimensional material are separated by a vacuum gap of size $d$. The nanoribbons are one-atom-thick with thicknesses, $t$, of 3.35 Å and 4.65 Å for graphene and silicene, respectively. The width of the nanoribbons is $W$, and their temperatures is fixed at $T$. Both graphene and silicene have hexagonal honeycomb structure. The lattice structure and vectors for graphene and silicene are shown in Fig. 1b. While graphene is completely flat, silicene has a buckled structure with a buckling of 0.45 Å (see Fig. 1b). The phonon conductance of the gap between the two nanoribbons is to be calculated.

## III. Methods

The phonon conductance of the vacuum gap, $\sigma$, due to a small temperature difference, $\delta T$, across the gap can be written as:

$$\sigma = \frac{1}{2\pi} \int_0^{\omega_c} \sigma_\omega d\omega \tag{1}$$

where $\omega$ is the angular frequency of phonons, $\omega_c$ is the cutoff frequency of phonons, and $\sigma_\omega$ is the spectral conductance found using the Landauer formula as [40]:

$$\sigma_\omega = \frac{\hbar\omega}{A} \frac{\partial f(\omega,T)}{\partial T} \Xi(\omega) \tag{2}$$

In Eq. 2, $A$ is the cross sectional area of the gap perpendicular to the transport direction, $\hbar$ is the reduced Planck constant, $f = \left(e^{\frac{\hbar\omega}{k_B T}} - 1\right)^{-1}$ is the Bose-Einstein distribution function of phonons with $k_B$ being the Boltzmann constant, and $\Xi$ is the transmission function of phonons across the gap. The transmission function of phonons is found using a three-dimensional atomistic Green's function (AGF) method. In this method, the system is divided into three regions, namely the left contact (or left lead), device, and right contact (or right lead) regions as schematically shown in Fig. 1a. The device region contains the vacuum gap as well as a few layers of carbon or silicon atoms at each side. Since the device region is very small, a ballistic phonon transport can be assumed. The transmission function can be found using the Caroli's formula as [41]:

$$\Xi(\omega) = \text{Tr}(\Gamma_L G_d^\dagger \Gamma_R G_d) \tag{3}$$

where Tr and the superscript † indicate the trace and Hermitian operators, respectively, and $G_d$ is the subset of the overall Green's function of the system which is associated with the device region. The device subset of the Green's function, $G_d$, can be found as [42]:

$$G_d = [\omega^2 I - H_d - \Sigma]^{-1} \tag{4}$$

where I is the unit dyad and $H_d$ is the device subset of the dynamical matrix of the system, $H$, given by [40,42]:

$$H_{ij} = \frac{1}{\sqrt{M_i M_j}} \begin{cases} \frac{\partial^2 U}{\partial u_i \partial u_j} & , i \neq j \\ -\sum_{i \neq j} \frac{\partial^2 U}{\partial u_i \partial u_j} & , i = j \end{cases} \quad (5)$$

In Eq. 5, $U$ is the potential energy function of the system, $u_i$ and $u_j$ are the spatial displacements associated with degrees of freedom $i$ and $j$, respectively, and $M_i$ and $M_j$ are the atomic masses corresponding to degrees of freedom $i$ and $j$, respectively. In Eq. 4, $\Sigma = \Sigma_L + \Sigma_R$, where $\Sigma_L$ and $\Sigma_R$ are the self-energy matrices of the left and right contacts, respectively, and they are found as [40]:

$$\Sigma_L = \tau_L g_L \tau_L^\dagger \quad (6a)$$

$$\Sigma_R = \tau_R g_R \tau_R^\dagger \quad (6b)$$

where $\tau_L$ and $\tau_R$ are the subsets of the dynamical matrix connecting the left and right contacts to the device region, respectively, and $g_L$ and $g_R$ are the uncoupled Green's function of the left and right contacts, respectively, given by:

$$g_L = [(\omega^2 I + i0^+) - H_L] \quad (7a)$$

$$g_R = [(\omega^2 I + i0^+) - H_R] \quad (7b)$$

In Eqs. 7a and 7b, $H_L$ ($H_R$) is the subset of dynamical matrix associated with the left (right) contact, and $0^+ = \delta_0 \omega^2$ with $\delta_0$ being an infinitesimally small positive number accounting for the phonon exchange between the contacts and the surroundings [42]. In our simulations, $\delta_0 = 10^{-4}$ is assumed as reducing $\delta_0$ below $10^{-4}$ does not affect the simulation results. The terms $\Gamma_L$ and $\Gamma_R$ in Eq. 3 are found using the self-energy matrices of the left and right contacts as $\Gamma_{\rho=L,R} = i(\Sigma_\rho - \Sigma_\rho^\dagger)$.

The Tersoff potential is used for modeling the interatomic interactions in graphene [43] and silicene [44] nanoribbons, while the Lennard-Jones potential [45,46] is used for modeling the interactions between atoms separated by the vacuum gap. Since the device region required for convergence of the AGF simulations is small, the Lennard-Jones potential is computed between all the atoms located in this region.

To verify the accuracy of our implementation of the AGF method, we have compared our results obtained for phonon conductance or transmission function of three different systems with those published in literature. Figure 2a compares the results of the current study for the conductance of a one-dimensional chain of silicon atoms multiplied by the surface area of the chain, $\sigma \cdot A$, with those obtained by Zhang et al. [47] at various temperatures. The strength of the Si-Si bonds is assumed to be 32 N/m. Figure 2b shows the spectral thermal conductance of an infinitely long graphene nanoribbon of width 10 nm at a temperature of $T = 300$ K. The Tersoff potential has been used for modeling the interatomic forces in graphene [43]. The results from the current study have been compared with the work by Li et al. [48]. Finally, the transmission function obtained for a vacuum gap between one-dimensional atomic chains of silicon and platinum under a bias voltage of 0.8 V is compared with those by Jarzembski et al. [21] in Figure 2c. The Si-Si and Pt-Pt bond strengths are assumed as 6.16 N/m and 6.31 N/m, respectively, while the Lennard-Jones potential and the Coulomb force have been used for modeling the van der Waals and electrostatic forces through the vacuum gap, respectively [21]. The agreement of the results of the current study with the previous work verifies the accuracy of the implemented AGF method for studying the thermal conductance of a vacuum gap between two nanoribbons of two-dimensional materials.

## IV.    Results and Discussion

Figure 3 shows the phonon conductance, $\sigma$, of a gap between two graphene nanoribbons with a width of $W = 10.08$ nm versus temperature for three gap sizes of $d = 0$, 0.2, and 1 Å. The phonon conductance for a gap of size $d = 0$ corresponds to thermal conductance of a single nanoribbon, which is 4.3 GW/(m$^2$.K) at room temperature and is in agreement with previous studies [48-50].

The gap size, $d$, is defined as the displacement of one of the nanoribbons by distance $d$ relative to the case where the two nanoribbons are connected forming a continuous nanoribbon (see the inset of Fig. 3). It is seen that thermal conductance of the gap drops by more than one order of magnitude when a gap as small as 0.2 Å exists between the two nanoribbons. The conductance continues to decrease with increasing the gap to 1 Å. This reduction in thermal conductance with increasing $d$ can be explained by considering that the phonon transport across the gap is mediated by the interatomic interactions across the gap which are modeled using the Lennard-Jones potential. The Lennard-Jones potential between atoms $i$ and $j$ across the gap, $V_{ij}$, is inversely proportional to the interatomic distance, $r_{ij}$, as $V_{ij} = 4\varepsilon \left[ \left( \frac{x}{r_{ij}} \right)^{12} - \left( \frac{x}{r_{ij}} \right)^{6} \right]$, where $\varepsilon$ is the potential depth and $x$ is the distance at which the potential energy between the two atoms is zero. As the distance between atoms increases with increasing the gap size, the interatomic potential rapidly decreases resulting in a sharp drop in thermal conductivity (see Fig. 5b for $V_{ij}$ between two closest carbon atoms located across the gap from each other). It is also seen that thermal conductance of the vacuum gap initially increases with increasing the temperature and then eventually plateaus. For ballistic phonon transmission across a sub-nanometric gap, the transmission function, $\Xi$, predicted using the atomistic Green's function is independent of the temperature, such that thermal conductance determined using Eqs. 1 and 2 has similar temperature dependence as the specific heat of the material, $c$ ($c \propto \int_0^\infty \hbar \omega g(\omega) \frac{\partial f(\omega,T)}{\partial T} d\omega$, where $g$ is the density of states). Initially, the number of phonons with a given frequency $\omega$ (which has a temperature-dependence as $\frac{\partial f(\omega,T)}{\partial T}$) sharply increases with increasing the temperature, and then it plateaus at the Debye temperature (see the inset of Fig. 3 for $\hbar \omega \frac{\partial f(\omega,T)}{\partial T}$ versus temperature). Figure 4a shows the spectral distribution of phonon conductance at three temperatures of 200 K, 250 K, and 300 K for a vacuum gap of size 0.2 Å between graphene nanoribbons. The dispersion relation of phonons for a graphene sheet, obtained using the Tersoff interatomic potential, is shown in Fig. 4b [30]. As will be discussed later, the phonon conductance of the nanoribbons with $W > 10$ nm varies very slightly with increasing the width. As such, the dispersion relation of a graphene sheet can be used for the nanoribbons with an acceptable accuracy. The Γ-M region of the dispersion relation is relevant to this study (see the inset of Fig. 4b). Figure 4b shows that the phonon conductance at low frequencies ($\omega/(2\pi) < 24$ THz) is solely driven by acoustic phonons. Conversely, the optical phonons mediate phonon conductance at large frequencies ($\omega/(2\pi) > \sim 39$ THz). The peak spectral conductance occurs around a frequency of ~ 29 THz, where the phonon conductance is contributed by the longitudinal acoustic (LA) and out-of-plane optical (ZO) phonons.

Figures. 4a and 4b demonstrate that the acoustic phonons have the dominant contribution to phonon conductance for a vacuum gap of $d = 0.2$ Å at the three considered temperatures. It is also seen from Fig. 4a that the contribution of high-frequency optical phonons (LO and TO phonons) to the conductance of the gap increases from 2.3% to 6.8% and 12.2% when the temperature increases from 200 K to 250 K and 300 K, respectively. This observation can be explained by considering the fact that the enhancement of the mean energy of phonons, given by $\hbar\omega \frac{\partial f(\omega,T)}{\partial T}$, with increasing the temperature is more significant at larger frequencies where the LO and TO phonons are spectrally located (See the inset of Fig. 4a).

The total thermal conductance of a gap between two graphene nanoribbons with a width of $W = 10.08$ nm at a temperature of $T = 300$ K is shown versus the size of the gap $d$ in Fig. 5a. Three different regimes for phonon conductance can be realized in Fig. 5a. In these regimes, the phonon conductance decays exponentially with the gap size. When $d < 1.62$ Å, thermal conductance rapidly reduces with the gap size as $\sigma \propto e^{-10.0d}$. For 1.62 Å $< d <$ 6 Å, the phonon conductance continues to decrease with increasing the gap size. However, the reduction rate reduces significantly, and the phonon conductance follows an exponential decay with the gap size as $\sigma \propto e^{-2.0d}$. When $d > 6$ Å, the reduction rate of the phonon conductance further reduces as $\sigma \propto e^{-1.3d}$. The existence of three different regimes for phonon conductance versus gap size can be explained by considering the Lennard-Jones potential which is used for modeling the interatomic interactions across the vacuum gap. The absolute value of the Lennard-Jones potential between two closest atoms across the gap, $|V_{ij}|$, versus the vacuum gap distance, $d$, is shown in Fig. 5b. As shown in the inset of Fig. 5b, the distance between the two closest atoms, $r$, is related to the vacuum gap size, $d$, the C-C bond length ($l_{C-C} = 1.42$ Å), and the C-C bond angle ($\alpha_{C-C} = 120°$) as $r^2 = l_{C-C}^2 + d^2 - 2l_{C-C}d\cos(150°)$. Figure 5b shows that when $d < 1.62$ Å, the repulsive forces in the Lennard-Jones potential (which are due to the overlap of electronic orbitals), completely dominates the total force between the two atoms resulting in a gap dependence of $\sigma \propto e^{-10.0d}$ for the conductance. As the distance between the two atoms increases over $d = 1.62$ Å, the repulsive forces continuously decrease while the attractive forces become greater. When the gap size is in the range of 1.62 Å $< d <$ 6 Å, both the attractive and repulsive forces have significant contributions to the Lennard-Jones potential. In this regime, as the net force between the atoms reduces, the reduction rate of the conductance with the gap size decreases to $\sigma \propto e^{-2.0d}$. As the gap size further increases to above 6 Å, the attractive van der Waals force dominates the interatomic interactions, and the conductance reduces exponentially with distance as $\sigma \propto e^{-1.3d}$. The conductance also closely follows a gap-size dependence of $\sigma \propto d^{-10.4}$ when $d > 6$ Å, which is in agreement with a previous study on the conductance of a vacuum gap with a size greater than 5 Å between two silicon plates performed using the density functional theory for modeling the interatomic potentials [20]. It should be also mentioned that the same three exponential regimes can be found for nanoribbons with smaller widths of $W = 5.40$ and $1.99$ nm.

The spectrum of the phonon conductance of the vacuum gap between graphene nanoribbons is strongly dependent on the size of the gap. The spectral phonon conductance is plotted in Figs. 6a-6c for three gap sizes of $d = 0.2$, $2.5$, and $10$ Å, for which the conductance is driven by the repulsive, repulsive-attractive, and attractive forces, respectively. The graphene nanoribbons have a width of $W = 10.08$ nm and a temperature of $T = 300$ K. Both acoustic and optical phonons contribute to phonon conductance at the smallest gap size of $d = 0.2$ Å, where the repulsive forces have dominant contribution on phonon conductance. However, the low-frequency acoustic phonons become the dominant contributor to the conductance at larger gap sizes of $d = 2.5$ and $10$ Å, for which the contribution from attractive forces to conductance is non-negligible or dominant. The low-frequency acoustic phonons have a longer wavelength, and thus the probability of their transmission through a large vacuum gap is greater than that of high-frequency acoustic phonons with a shorter wavelength. For the same reason, the contribution of LO and TO phonons, which are active at large frequencies and thus have a small wavelength, diminishes when the gap size increases to $d = 2.5$ and $10$ Å.

The phonon conductance of a gap of size $d = 0.2$ Å between two graphene nanoribbons at a temperature of $T = 300$ K is shown versus the nanoribbon width $W$ in Fig. 7. The phonon conductance of the gap initially slightly decreases from ~$0.488$ GWm$^{-2}$K$^{-1}$ for $W = 1$ nm to ~$0.431$ GWm$^{-2}$K$^{-1}$ for $W = 10$ nm (by 11.7%). Increasing the nanoribbon width beyond $W = 10$ nm does not change the phonon conductance of the gap appreciably. Since the phonon conductance remains almost constant as $W$ increases, the phonon conductance through the total surface area of the gap (i.e., $\sigma \cdot A$, where $A = W \cdot t$ and $t = 0.335$ nm is the thickness of the graphene nanoribbon) linearly increases with $W$ (see the inset of Fig. 7).

Lastly, we study the phonon conductance of a vacuum gap between two silicene nanoribbons. The total conductance for silicene nanoribbons with a width, $W$, of 10.56 nm at a temperature of $T = 300$ K is shown versus the gap size in Fig. 8a. Similar to the case of graphene nanoribbons, the gap conductance between silicene nanoribbons exponentially decays with increasing the gap size. The Lennard-Jones potential between the two nearest silicon atoms across the gap versus the gap size is shown in Fig. 8b. Since the bond length of the silicon atoms in silicene (2.15 Å) is much greater than the carbon atoms in graphene (1.42 Å), the repulsive forces between the silicon atoms do not dominate the Lennard-Jones potential even for a gap size as small as $d = 0.2$ Å. Indeed, the attractive term in the Lennard-Jones potential is 1.7 times greater than the repulsive term when $d = 0.2$ Å. When $d < 2.5$ Å, the Lennard-Jones potential has contributions from both repulsive and attractive forces. In this regime, the conductance of the gap reduces with the gap size as $e^{-2.5d}$, which is approximately the same as that found for graphene nanoribbons in the regime where both attractive and repulsive forces drive the gap conductance ($\sigma \propto e^{-2.0d}$). When $d > 2.5$ Å, the attractive forces dominate the Lennard-Jones potential between silicon atoms. In this regime, the gap conductance follows an exponential decay as $e^{-1.3d}$, which is the same as the one found for graphene in the attractive-force regime. Additionally, when $d > 5$ Å, $\sigma$ for silicene nanoribbons follows a power law as $\sigma \propto d^{-9.6}$, which is consistent with that found for a gap between graphene nanoribbons and the one reported in the literature [20].

Figures 8a also compares the phonon conductance of the gap between two silicene nanoribbons with that between two graphene nanoribbons with approximately the same width ($W = 10.56$ and 10.08 nm for silicene and graphene nanoribbons, respectively). A temperature of $T = 300$ K is assumed for the nanoribbons. When $d = 0.2$ Å, the phonon conductance of the gap between silicene nanoribbons is more than two orders of magnitude smaller than that between graphene nanoribbons. As $d$ increases, the gap conductance for both graphene and silicene nanoribbons decreases exponentially. However, the rate of reduction for the gap between graphene nanoribbons ($e^{-10.0d}$) is much greater than silicene nanoribbons ($e^{-2.5d}$), such that the gap conductance for silicene overtakes that for graphene when $d > 0.8$ Å. The phonon conductance of the gap depends on the Lennard-Jones potential between atoms across the gap as well as the Tersoff potential between atoms in each nanoribbon. The Tersoff potential between carbon atoms in graphene is significantly greater than that between silicon atoms in silicene. The Lennard-Jones potential, however, depends on the size of the vacuum gap as shown in Fig. 8b. When $d = 0.2$ Å, the Lennard-Jones potential for graphene is significantly greater than for silicene, and so is the gap conductance. As the gap increases, the Lennard-Jones potential between carbon atoms decreases very rapidly, and it falls below the one for silicon atoms at a gap size of $d = 0.5$ Å. Since the Lennard-Jones potential for graphene decreases more rapidly with increasing $d$ than for silicene, $\sigma$ for graphene nanoribbons falls below $\sigma$ for silicene nanoribbons when $d$ exceeds 0.8 Å.

The spectral conductance of a gap between two silicene nanoribbons at a temperature of $T = 300$ K is shown in Fig. 8c and 8d for gap sizes of $d = 0.2$ and 5 Å, respectively. The conductance of the gap between silicene nanoribbons is driven by both repulsive and attractive forces when $d = 0.2$ Å, while the attractive forces have the dominant contribution to the conductance when $d = 5$ Å. The dispersion relation of phonons for a silicene sheet [51] is also shown in the inset of Fig. 8c. Figures 8b and 8c show that, similar to the case of graphene nanoribbons, the phonon conductance in the repulsive-attractive and attractive regimes is dominated by the contribution from acoustic phonons which are located at frequencies smaller than 4 THz.

## V. Conclusions

The phonon conductance in the extreme near-field regime was theoretically studied for a vacuum gap between two semi-infinite, in-plane nanoribbons of two-dimensional materials (graphene and silicene). The conductance of the gap was modeled using the atomistic Green's function approach and by employing the Tersoff and Lennard-Jones interatomic potentials. It was found that even a small gap of size $d = 0.2$ Å between the two nanoribbons can significantly impede phonon transfer and causes a more than one order of magnitude drop in phonon conductance compared to the case of connected nanoribbons ($d = 0$). It was shown that for both silicene and graphene, the phonon conductance follows exponential decays with increasing the size of the gap. The conductance of the gap between graphene nanoribbons decreases with increasing the gap size, $d$, as $e^{-10.0d}$, $e^{-2.0d}$, and $e^{-1.3d}$ in the regimes where the repulsive forces, the combined repulsive and attractive forces, and the attractive forces drive the interatomic potential across the gap, respectively. Due to a large bond length for silicon atoms, the repulsive forces between silicon atoms across the gap are weak, and the conductance of the gap between silicene nanoribbons follows mild exponential decay with the gap size as $e^{-2.5d}$ and $e^{-1.3d}$ when $d < 2.5$ Å (repulsive-attractive regime) and $d > 2.5$ Å (attractive regime), respectively. The phonon conductance of the gap is mostly mediated by acoustic phonons, except for the case of small gap sizes between graphene nanoribbons, where the repulsive forces are dominant. In this case, there is a non-negligible contribution from optical phonons to the total conductance. It was also found that the contribution from low-frequency acoustic phonons to the conductance of the gap between graphene nanoribbons dominates at larger gap sizes as these phonons have a larger wavelength compared to high-frequency acoustic phonons. Lastly, it was found that the phonon conductance of the gap does not vary strongly with the width of the nanoribbons, such that the conductance for the total cross-sectional area of the gap increases linearly with the width.

**Acknowledgements**

This work is supported by the National Science Foundation under Grant No. CBET-2046630. The authors thank Dr. Zheyong Fan for helping with the Atomistic Greens Function method.

[25] Viloria, Mauricio Gómez, Yangyu Guo, Samy Merabia, Riccardo Messina, and Philippe Ben-Abdallah. "Radiative heat exchange driven by the acoustic vibration modes between two solids at the atomic scale." *arXiv preprint arXiv:2302.00520* (2023).

[26] Kittel, Achim, Wolfgang Müller-Hirsch, Jürgen Parisi, Svend-Age Biehs, Daniel Reddig, and Martin Holthaus. "Near-field heat transfer in a scanning thermal microscope." *Physical review letters* 95, no. 22 (2005): 224301.

[27] Kim, Kyeongtae, Bai Song, Víctor Fernández-Hurtado, Woochul Lee, Wonho Jeong, Longji Cui, Dakotah Thompson et al. "Radiative heat transfer in the extreme near field." *Nature* 528, no. 7582 (2015): 387-391.

[28] Kloppstech, Konstantin, Nils Könne, Svend-Age Biehs, Alejandro W. Rodriguez, Ludwig Worbes, David Hellmann, and Achim Kittel. "Giant near-field mediated heat flux at the nanometer scale." *arXiv preprint arXiv:1510.06311* (2015).

[29] Cui, Longji, Wonho Jeong, Víctor Fernández-Hurtado, Johannes Feist, Francisco J. García-Vidal, Juan Carlos Cuevas, Edgar Meyhofer, and Pramod Reddy. "Study of radiative heat transfer in Ångström-and nanometre-sized gaps." *Nature communications* 8, no. 1 (2017): 14479.

[30] Feng, Tianli, and Xiulin Ruan. "Four-phonon scattering reduces intrinsic thermal conductivity of graphene and the contributions from flexural phonons." *Physical Review B* 97, no. 4 (2018): 045202.

[31] Cheng, Yajuan, Zheyong Fan, Tao Zhang, Masahiro Nomura, Sebastian Volz, Guimei Zhu, Baowen Li, and Shiyun Xiong. "Magic angle in thermal conductivity of twisted bilayer graphene." *Materials Today Physics* 35 (2023): 101093.

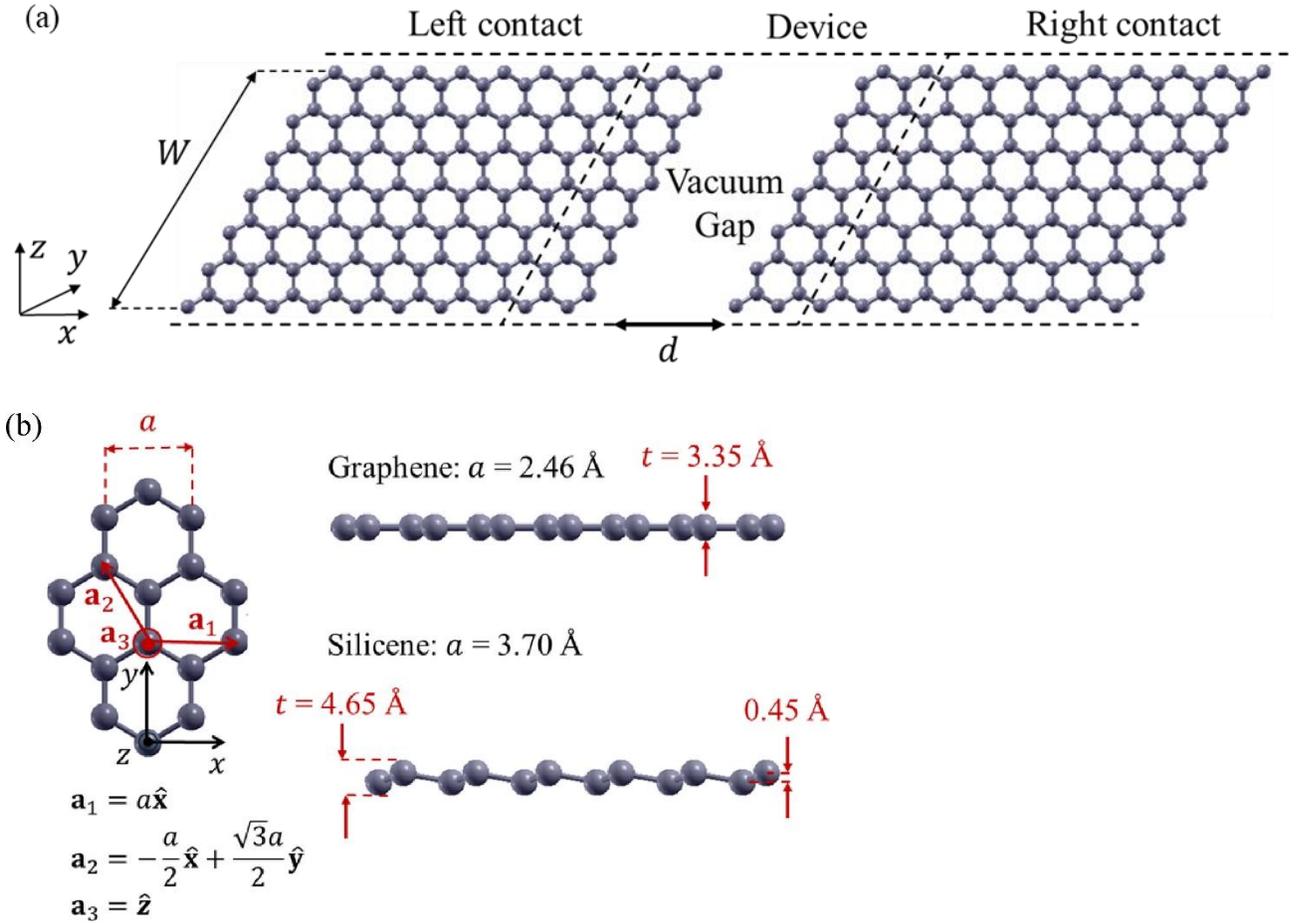

**Figure 1** – (a) A schematic of the system under study. Two in-plane, semi-infinite nanoribbons of graphene or silicene are separated by a vacuum gap of size $d$. The nanoribbons have a width $W$, and they are at temperature $T$. The phonon conductance of the gap is to be calculated using the atomistic Green's function (AGF) method. In the AGF method, the system under study is divided into a left contact, a device, and a right contact region. (b) The lattice structure, lattice vectors, and a side view of the graphene and silicene nanoribbons.

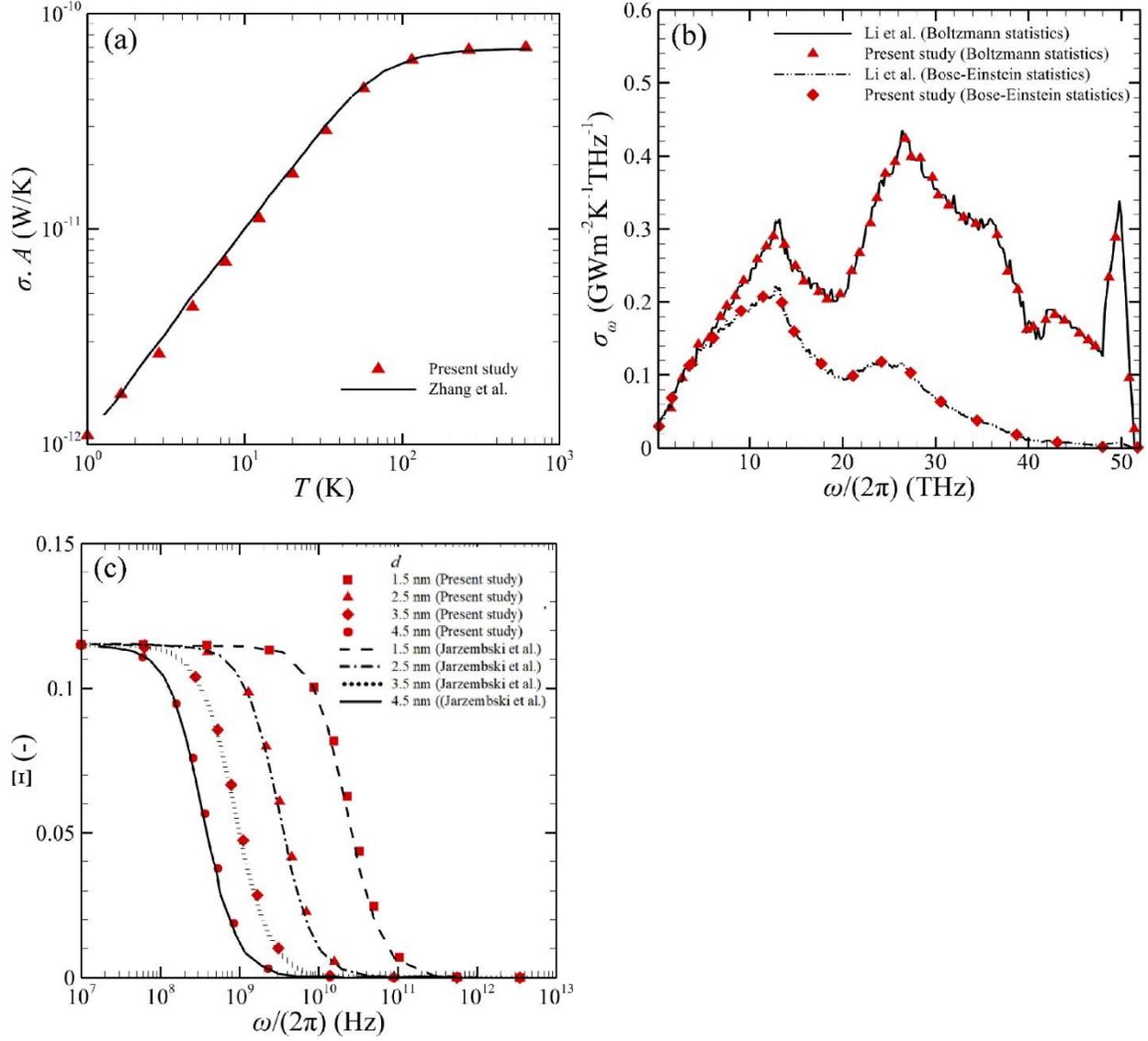

**Figure 2** - Verification of the implemented atomistic Green's function method for modeling the phonon conductance of a vacuum gap between two semi-infinite media. (a) The phonon conductance, $\sigma$, multiplied by the surface area, $A$, for a one-dimensional chain of silicon atoms. The results obtained in the present study are compared with those from Zhang et al. [47]. (b) The spectral conductance, $\sigma_\omega$, of a graphene nanoribbon with a width of $W = 10$ nm and a temperature of $T = 300$ K obtained in the present study in comparison to that from Li et al. [48]. (c) The transmission function of phonons through a vacuum gap of size $d$ between a one-dimensional chain of silicon atoms and a one-dimensional chain of platinum atoms. The results of the current study are compared with those from Jarzembski et al. [21].

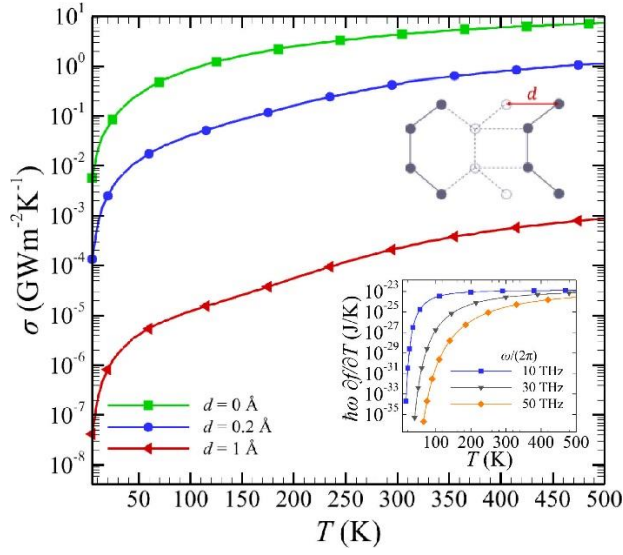

**Figure 3** – Phonon conductance, $\sigma$, of a vacuum gap of size $d$ between two graphene nanoribbons with a width of $W = 10.08$ nm versus temperature, $T$. The inset of the figure shows a schematic depicting the vacuum gap between the two nanoribbons as well as the mean energy of phonons, $\hbar\omega \frac{\partial f(\omega,T)}{\partial T}$, versus temperature, $T$, for three frequencies of $\frac{\omega}{2\pi} = 10$, 30, and 50 THz.

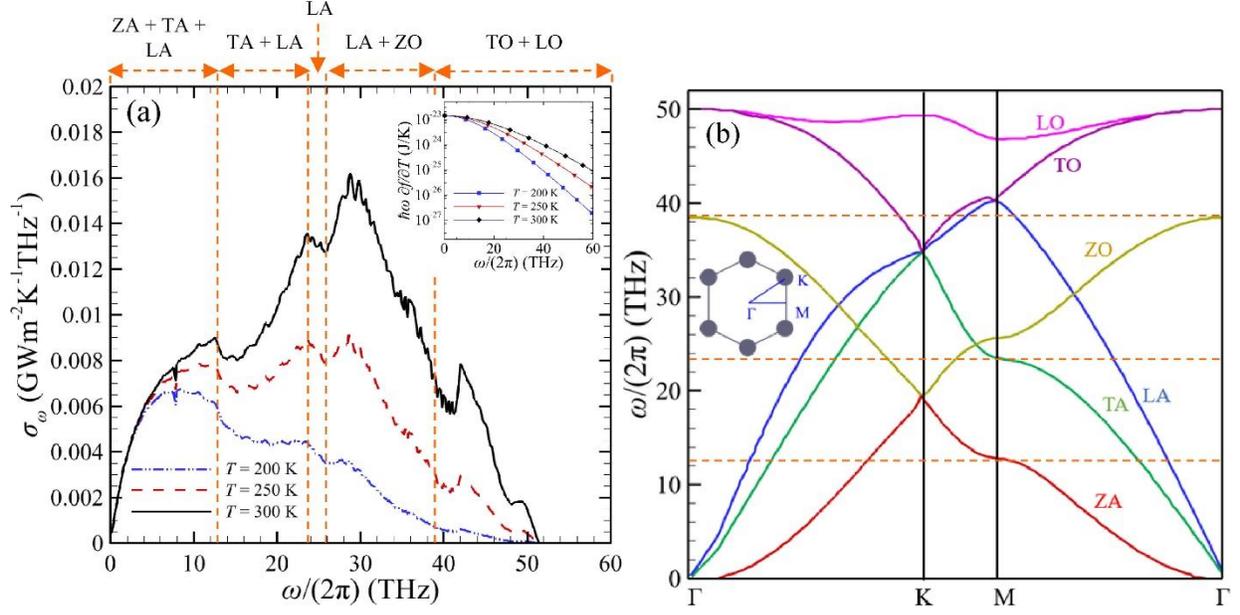

**Figure 4** – (a) Spectral phonon conductance, $\sigma_\omega$, of a vacuum gap of size $d = 0.2$ Å between two graphene nanoribbons with a width of $W = 10.08$ nm at three temperatures of $T = 200$ K, 250 K, and 300 K. The inset of the figure shows the mean energy of phonons, $\hbar\omega\frac{\partial f(\omega,T)}{\partial T}$, versus frequency, $\frac{\omega}{2\pi}$, for three temperatures of $T = 200, 250,$ and 300 K. (b) The dispersion relation of phonons for a graphene sheet as obtained using the Tersoff interatomic potential [30]. LA, TA, ZA, LO, TO, and ZO stand for longitudinal acoustic, transverse acoustic, out-of-plane acoustic, longitudinal optical, transverse optical, and out-of-plane optical phonons, respectively.

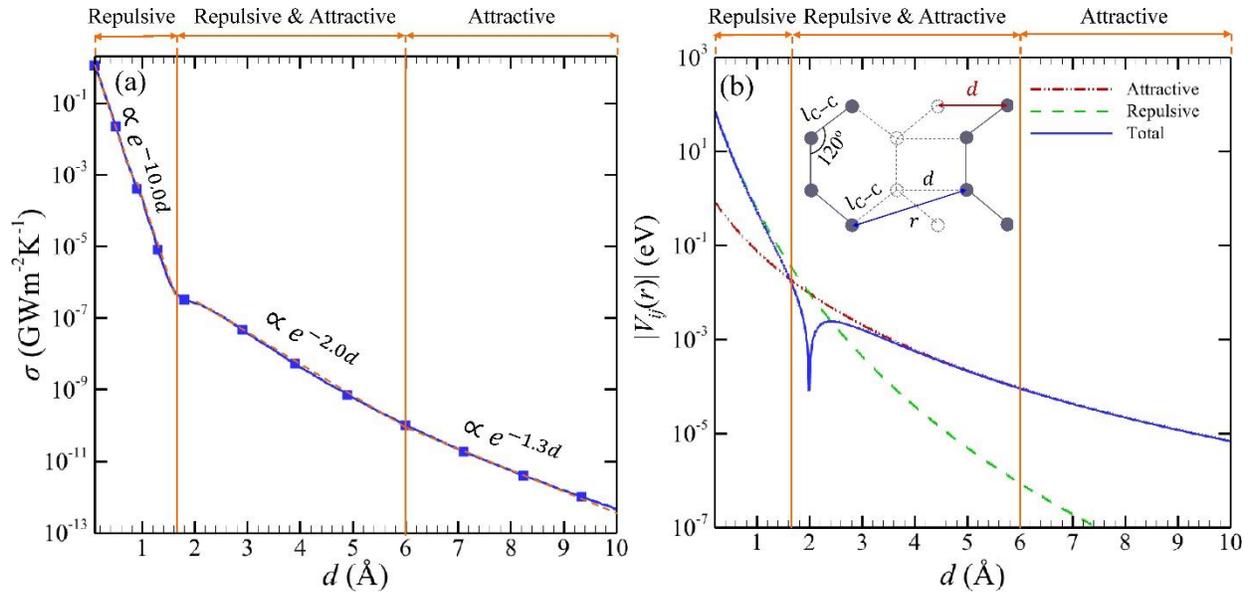

**Figure 5** – (a) The phonon conductance of a vacuum gap between two graphene nanoribbons with a width of $W = 10.08$ nm and a temperature of $T = 300$ K versus the size of the gap $d$. (b) The Lennard-Jones potential function, $V_{ij}$, between two closest carbon atoms across the vacuum gap versus the gap size, $d$.

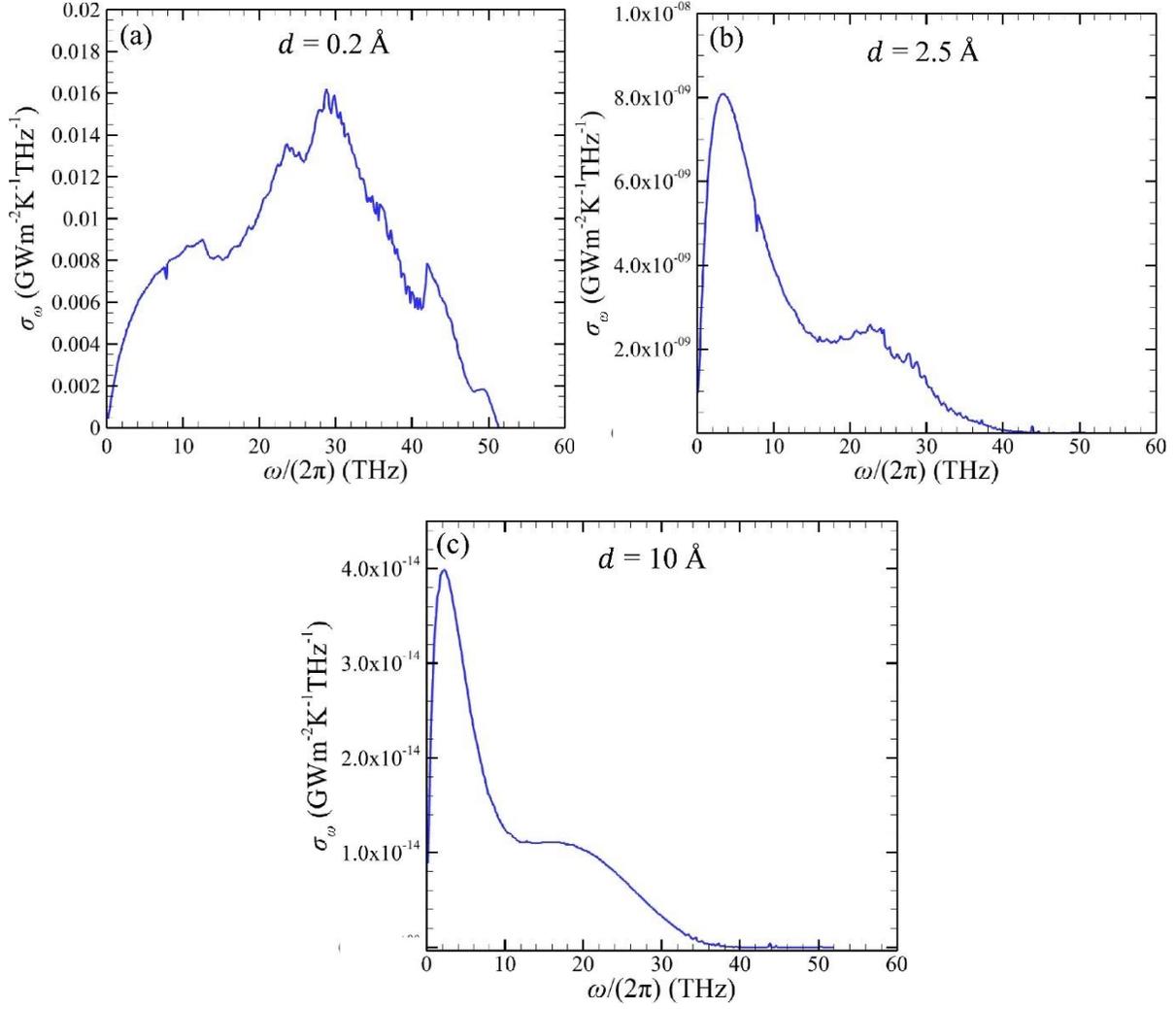

**Figure 6** – The spectral phonon conductance, $\sigma_\omega$, of a vacuum gap with size $d$ between two graphene nanoribbons with a width of $W = 10.08$ nm and a temperature of $T = 300$ K. (a) $d = 0.2$ Å, (b) $d = 2.5$ Å, and (c) $d = 10$ Å.

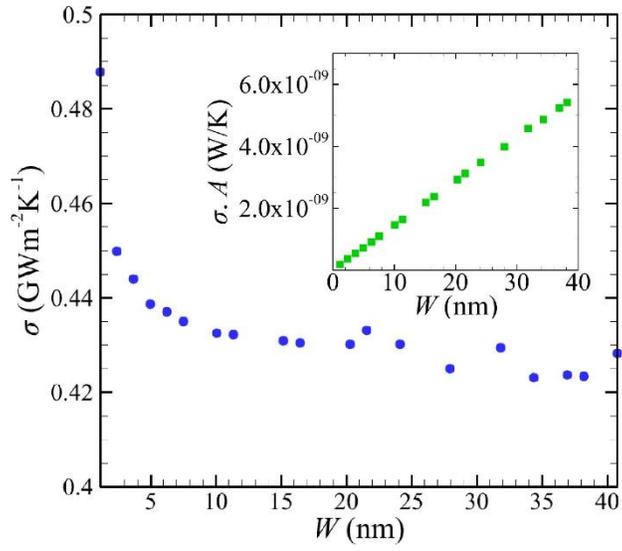

**Figure 7** – The phonon conductance of a vacuum gap of size $d = 0.2$ Å between two graphene nanoribbons with a width of $W$ and a temperature of $T = 300$ K. The inset shows the phonon conductance multiplied by the surface area of the gap, $\sigma \cdot A$.

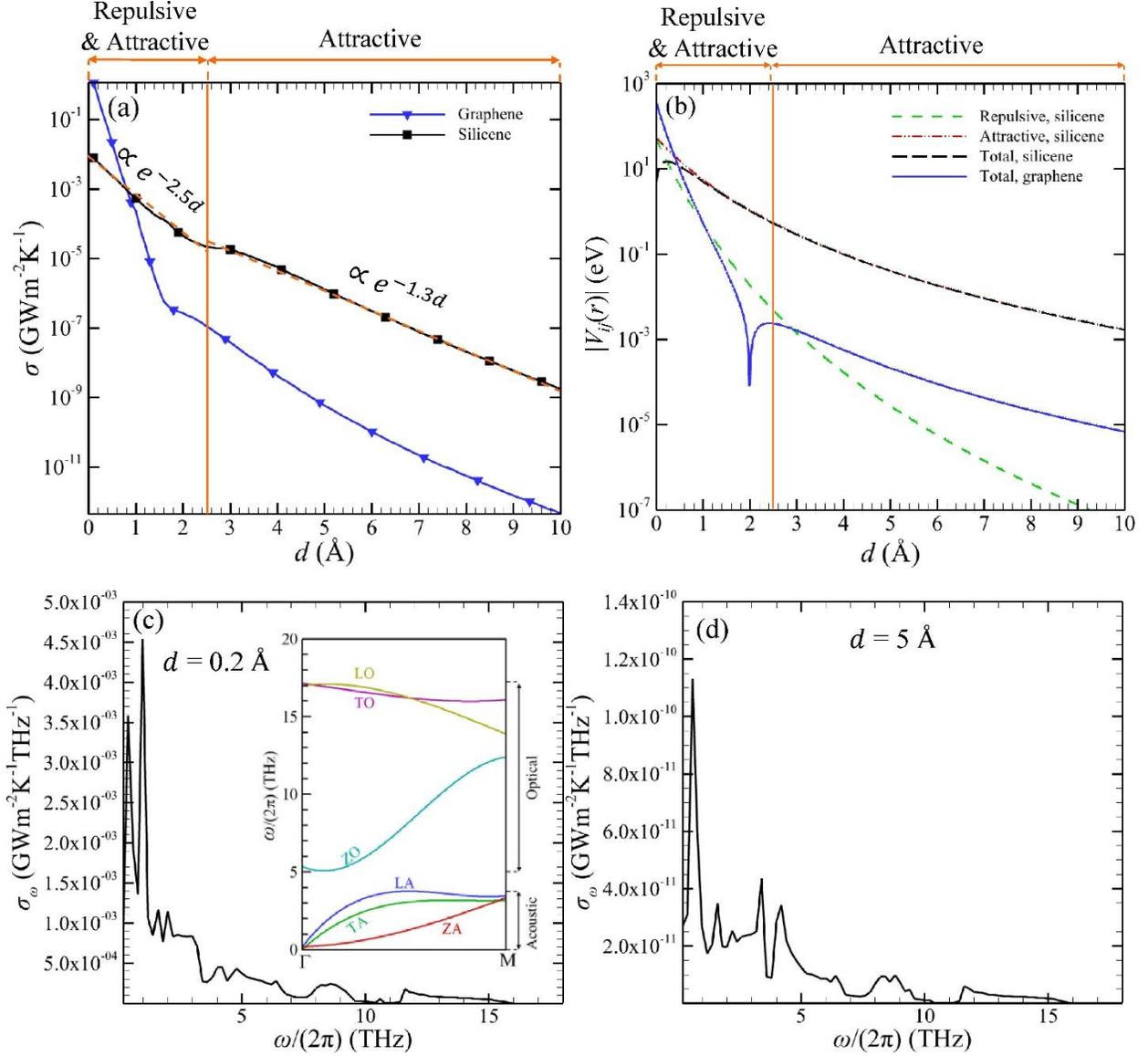

**Figure 8** – (a) The total phonon conductance of a gap between two silicene nanoribbons with a width of $W = 10.56$ nm versus the one between two graphene nanoribbons with a width of $W = 10.08$ nm as a function of the gap size $d$. A temperature of $T = 300$ K is assumed. (b) The Lennard-Jones potential function, $V_{ij}$, between two closest silicon atoms across the vacuum gap between silicene nanoribbons versus the gap size, $d$. (c, d) The spectral phonon conductance of vacuum gaps with sizes of (c) $d = 0.2$ Å and (d) $d = 5$ Å between two silicene nanoribbons with a width of $W = 10.56$ nm and a temperature of $T = 300$ K. The dispersion relation of a silicene sheet [51] is shown in the inset of Pabel (c).